\documentclass[12pt]{article}

\usepackage{arxiv}

\usepackage[utf8]{inputenc} 
\usepackage[T1]{fontenc}    
\usepackage[hidelinks]{hyperref}       
\usepackage{url}            
\usepackage{booktabs}       
\usepackage{amsfonts}       
\usepackage{nicefrac}       
\usepackage{microtype}      
\usepackage{lipsum}
\usepackage{graphicx}
\usepackage{setspace}
\usepackage{natbib}

\graphicspath{ {./images/} }

\title{Collaboratively adding context to social media posts reduces the sharing of false news}

\author{
 Thomas Renault \\
  Université Paris 1 Panthéon-Sorbonne\\
  \texttt{thomas.renault@univ-paris1.fr} \\
   \And
 David Restrepo - Amariles \\
  HEC Paris\\
  \texttt{restrepo-amariles@hec.fr} \\
  \And
Aurore Troussel - Clément \\
HEC Paris\\
  \texttt{aurore.troussel@hec.edu} \\
}

\begin{document}
\maketitle
\begin{abstract}
We build a novel database of around 285,000 notes from the Twitter Community Notes program to analyze the causal influence of appending contextual information to potentially misleading posts on their dissemination. Employing a difference in difference design, our findings reveal that adding context below a tweet reduces the number of retweets by almost half. A significant, albeit smaller, effect is observed when focusing on the number of replies or quotes. Community Notes also increase by 80\% the probability that a tweet is deleted by its creator. The post-treatment impact is substantial, but the overall effect on tweet virality is contingent upon the timing of the contextual information's publication. Our research concludes that, although crowdsourced fact-checking is effective, its current speed may not be adequate to substantially reduce the dissemination of misleading information on social media.
\end{abstract}


\newpage

\onehalfspacing

\section{Introduction}

Addressing the proliferation of misinformation on social media is imperative due to its potential to erode public trust, compromise the integrity of information ecosystems, and undermine the democratic debate \citep{allcott2017social, grinberg2019fake}. Content moderation, which involves reviewing and monitoring user-generated content, is the primary strategy used by platforms to maintain user trust \citep{gillespie2018custodians} and is an obligation for very large online platforms under new regulations such as the European Union’s Digital Services Act (DSA). Determining methods of content moderation and evaluating their effectiveness is thus key in curtailing the spread of misinformation. 

Combating misinformation is challenging due to the rapid pace at which new content emerges, the inherent difficulties in discerning between genuine and deceptive information, and the need to strike a balance between upholding principles of free speech and preventing the spread of illegal or harmful content. Due to the complexities associated with content moderation, several platforms have introduced in recent years a form of crowdsourced moderation, by which users themselves report and flag content that may violate the platform's policies. Crowdsourced moderation complements the efforts of professional moderators and automated tools used by platforms.

In 2021, X (formerly Twitter) launched a pilot program in the United States to enable its users to propose contextual Notes for potentially false or misleading tweets. The Notes that receive enough positive ratings from a diverse set of users are then visible alongside tweets on the platform to provide other users with more information to make informed judgments about the content they encounter. This program has been praised by Elon Musk (X CEO) as largely increasing the trust and the reliability of the information published on X\footnote{ For example, in a tweet on October 6th, 2023, Elon Musk wrote that it is "so much harder to lie on this platform with @CommunityNotes keeping people honest”.}. However, their effectiveness is now being questioned by regulators. In October 2023, the European Commission opened a formal proceedings  to investigate whether X may have violated the DSA, specifically raising concerns about "the effectiveness of measures taken to combat information manipulation on the platform, notably the effectiveness of X's so-called Community Notes system in the EU [...]".\footnote{European Commission, Press release, “Commission opens formal proceedings against X under the Digital Services Act”, December 2023, and  Commission decision initiating proceedings pursuant to Article 66(1) of Regulation (EU) 2022/2065.}

In this paper, we shed light on this debate by quantifying the effect of Community Notes on the diffusion of false or misleading information on X. Our identification strategy combined with high-frequency information diffusion allow us to precisely estimate the causal effect of this content moderation method. The variation in tweet diffusion that we observe is only due to users' reaction to the context provided as X does not use Community Notes on its algorithm to automatically reduce the visibility of the tweet. We show that adding a Community Note approximately halves the number of shares (retweets) and increases the probability that a tweet will be deleted by its creator by 80\%. This post-treatment effect is significant, but the overall impact on the spread of false or misleading tweets is much more modest (-16\% to -21\%) due to the publication delay of the Notes.

\section{Literature}

Research shows that the spread of false information and harmful narratives can occur within minutes, gaining momentum and influencing public perception before corrective measures are taken \citep{vosoughi2018spread}.  Speedy response in content moderation is key to mitigate the harmful effects of false or misleading information as delays in moderating problematic content can allow it to reach a wider audience and contribute to the perpetuation of falsehoods \citep{friggeri2014rumor}. To reduce the reach of false news, one intervention approach involves enabling individuals exposed to such news to better assess its veracity \citep{lazer2018science}. Major platforms such as Facebook, Instagram and TikTok have adopted in recent years soft content moderation approaches that consist of applying a label or tag directly on content that appears to be false or misleading. Those warning labels indicate that an information is “disputed”, “false” or “unverified” after being checked by a third-party fact-checker. In theory, soft content moderation might be ineffective due to politically motivated reasoning, the risk of backfire effects, and the implied truth effect \citep{pennycook2020implied}. Nonetheless, despite these limitations, several experiments have shown that warning labels are broadly effective \citep{martel2023misinformation} and reduce the sharing of false news by 25\% \citep{mena2020cleaning} to 46\% \citep{pennycook2020implied}.

X's Community Notes system is different from the warning labels previously studied for at least three main reasons. First, it is not just a simple warning but a contextual piece of information explaining why, according to the author of the Note, the information in the tweet is misleading. It can be seen as a community-driven fact-checking system. On the effectiveness of fact-checking, \cite{henry2022checking} find in a randomized experiment that fact-checking reduces the sharing of false statements by about 45\%. Second, it is not generated by independent fact-checkers but by the crowd. In that regard, recent studies explore alternative methods for creating these labels, including using crowdsourced fact-checking to provide additional context or indicators. The results from this literature are mixed: while labels from expert fact-checkers tend to be more universally effective and align with established content moderation practices, labels generated through community-sourced fact-checking have also shown effectiveness \citep{allen2021scaling}.  Third, the design of the content moderation system has a large impact on user reaction. The choice of the words used (for example “disputed” versus “false”), the timing of the intervention (before, alongside or after the news) \citep{brashier2021timing}, or even just the size of the color of the text could affect the perceived veracity of those news. X uses the sentence “readers added context they thought people might want to know” before displaying the contextual Note, and little is known about the effectiveness of such design.

Most papers on the effectiveness of content moderation rely on experiments due to a lack of real data from the platform. To the best of our knowledge, only two papers in the literature directly analyze the effect of moderation on false news diffusion using data from platforms. The closest paper in the literature is by \cite{chuai2023roll} who find no evidence that the introduction of Community Notes significantly reduces engagement with misleading tweets on X. However, the results of this study do not allow for the measurement of the causal effect of Community Notes because their research design compares tweets with and without a Note (ex-post), rather than assessing the impact on the diffusion of a tweet as a consequence of attaching a Note. The only causal estimate in the literature is an internal A/B test at the launch of the program in the United States in 2021 \citep{wojcik2022birdwatch} on which the authors found a reduction in the number or retweets of 25 to 34\% when a Note is displayed under a tweet. However, these results should be interpreted with caution as the study was conducted internally at X, and the data used are not publicly available.

\section{Dataset}
\label{sec:headings}
\begin{figure} 
    \centering
    \caption{Example of a Community Note}
    \includegraphics{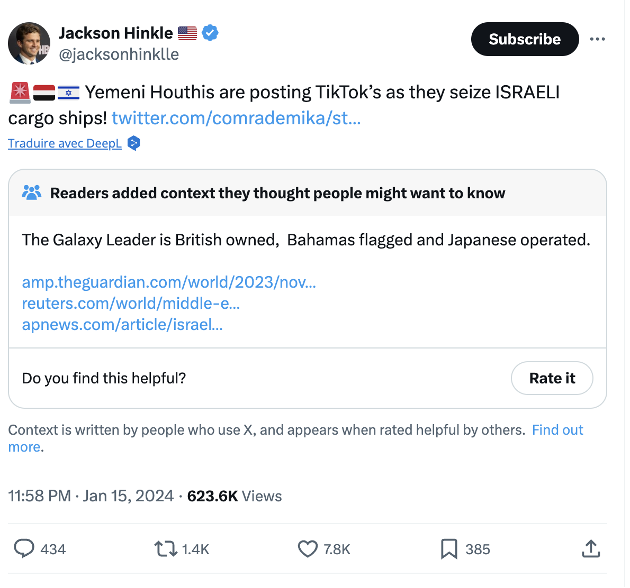}
    \label{fig:exnote}
\end{figure}

X's Community Notes system is a crowdsourced content moderation feature aimed at enabling users to propose and rate Notes providing contextual information to tweets. Notes are intended to provide context, fact-checking, or any relevant information that addresses the accuracy or veracity of the tweet. Community Notes are short-texts that should directly address the post’s claim in neutral or unbiased language and - when possible - cite high-quality sources. The system began as a pilot project in January 2021 under the name Birdwatch, and was rebranded as Community Notes in August 2021. Figure \ref{fig:exnote} shows an example of a Community Note: the block that starts with “Readers added context they thought people might want to know” contains the contextual information proposed by a Community Notes contributor and rated as Helpful by the Community. It is displayed below the tweet and before the buttons to comment, retweet, like or pin the misleading tweet.

Ratings from other contributors are used to determine note statuses (“Helpful”, “Not Helpful”, or “Needs More Ratings'')\footnote{To be able to write notes, new contributors must first pass a test phase during which they evaluate the notes written by other contributors. All users (1) with no recent notice of violations of X's Rules, (2) who joined X for more than 6 months and (3) with a verified phone number can join the Community Notes system and become a contributor. If their evaluations are deemed useful in rating these notes, they can start writing notes on the tweets of their choice.}. The algorithm used by X to compute a Note Helpfulness Score is open-source \citep{wojcik2022birdwatch} and Notes marking posts as "potentially misleading" with a score of 0.40 and above earn the status of Helpful and are displayed on posts. Note statuses aren't reached only by majority rule but also require agreement between contributors who have sometimes disagreed in their past ratings. This rule is used to diminish the risk of adversarial manipulations by requiring consensus across polarized groups. 

We download the Community Notes dataset, made publicly available by X for research purposes, which contains Community Notes that have been proposed, ratings on each Note from X users and Note Status History Data. From those dataset, we can distinguish Notes Currently Rated Helpful (CRH) from Notes that Needs More Ratings (NMR) and Notes Currently Rated Not Helpful (CRNH). We can also reconstruct - for CRH notes - the exact minute at which the note reached its Helpful status and thus became visible by everyone on the platform.

One limit of the public dataset provided by X is that it only contains the unique identifier of the tweet attached to each Note, but no information on the tweet content nor on the tweet virality. We thus use the Twitter API Pro to extract information about all tweets in our database (content, date of publication, language, username...). We also retrieve detailed diffusion information for each tweet, capturing minute-by-minute numbers of quotes, replies, and retweets. This allows us to analyze how tweets spread, and the dynamics of user engagement over time. This feature sets our research apart in the literature, enabling us to identify the causal effect of Community Notes on the diffusion of tweets (accessing this data requires a Twitter Pro API and costs \$5,000 per month).

\begin{figure} 
    \centering
    \caption{Number of notes and tweets}
    \includegraphics{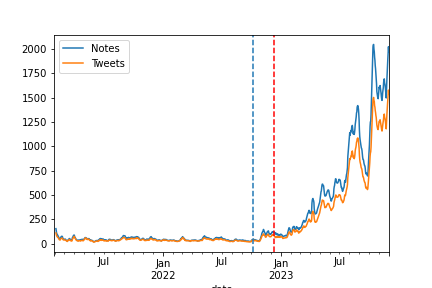}
    \begin{center}Note : This Figure shows the total number of Notes and the number of distinct tweets on our database (7-day rolling mean). The red vertical bar on December 11, 2022 represents the worldwide deployment of the system.\end{center}
     \label{fig:notetot}
\end{figure}

Our database is composed of a total of 286,198 Notes related to 220,641 distinct tweets between January 2021 and December 1st 2023. We can distinguish in Figure  \ref{fig:notetot} three periods : (1) the pilot phase period, from January 25, 2021 to March 3, 2022, (2) the progressive deployment of the program in the United States, with all US users seeing notes from October 6, 2022, and (3) the worldwide deployment of the system since December 11, 2022. We focus on the period between December 11, 2022, to December 1, 2023, to avoid having our results biased by the pilot phase or exclusive deployment in the US.

\begin{figure}[ht]
  \centering
  \begin{minipage}[b]{0.45\linewidth}
  \caption{Diffusion speed}
    \includegraphics[trim={1cm 0.5cm 1cm 0.5cm},clip,width=1\linewidth]{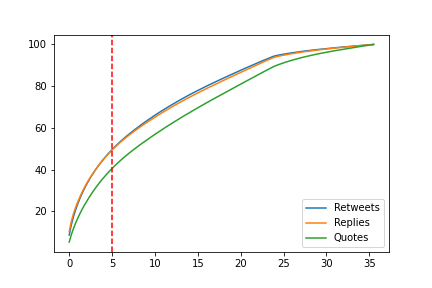}
    \begin{center}Note : This Figure shows the cumulative diffusion of retweets, replies and quotes per hour. The half-life for retweets and replies is represented with a vertical bar.\end{center}
    \label{fig:speed}
  \end{minipage}
  \hfill
  \begin{minipage}[b]{0.45\linewidth}
  \caption{Tweet deletion}
   \includegraphics[width=1\linewidth]{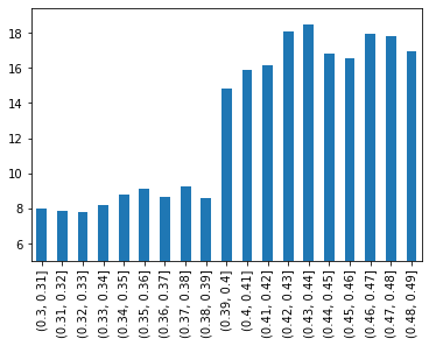}
    \begin{center}
    Note : This Figure shows the percentage of deleted tweets by Note Helpfulness Score\end{center}
    \label{fig:deletion}
  \end{minipage}
\end{figure}

Four interesting findings emerge from a first inspection of those dataset.  First, the diffusion process of tweets is extremely fast. When we focus on tweets with a Needs More Ratings status (i.e, tweets in our control group to avoid capturing the impact of moderation in the estimation), we find that around 50\% of the total number of retweets happens in the first 5 hours of the life of a tweet (80\% after 16 hours). Similar results are visible for replies while quotes are slightly slower (Figure \ref{fig:speed}). Second, on average, the delay between the publication of a misleading tweet and the publication of a contextual Note on X is around 15.5 hours (the median is 14.3 hours).  Although this delay has decreased since the launch of Community Notes, it remains significant relative to the speed at which information spreads on X. Third, most Notes are never published : only 11.3\% of Notes reach CRH status. This effect could be partially explained by the low quality of some proposed Notes, but also shows the difficulty of reaching a consensus (a Note Helpfulness score above 0.4) on controversial tweets with a crowdsourced approach. Fourth, when we analyze the percentage of deleted tweets by Note Helpfulness Score, we find that the percentage of deleted tweets is 80\% higher above the Helpfulness threshold of 0.4 than below the threshold (
15.8\% of tweets with Notes rated between 0.4 and 0.41 were deleted, compared to only 8.6\% for Notes rated between 0.38 and 0.39) (Figure \ref{fig:deletion})\footnote{The cut-off used by X is not exactly 0.4 and Notes with ratings falling between 0.39 and 0.40 may be marked as “Helpful” by X due to specific rounding criteria.}. Even if we cannot extract any other information about deleted tweets to push further this analysis - as due to X data management policies the content, timestamp, diffusion process before deletion or the name of the user are not available on deleted tweets - this result shows that Community Notes could also have a strong effect on misleading tweet visibility in increasing deletion of those tweets.

\section{Methodology}
\label{sec:headings}

Our framework is characterized by multiple time periods, variation in treatment timing, and staggered adoption (once a tweet is treated, it remains treated in the following periods). To estimate the average treatment effects in this difference in difference (DiD) setup, we use the \cite{callaway2021difference} estimators. For all tweets, we analyze the diffusion of quotes, replies, and retweets by analyzing the volume of virality per one hour interval from the publication of the tweet up to 36 hours after the publication. We use doubly robust DID estimators with bootstrapped standard errors and no anticipation. To maintain data integrity and minimize the influence of tweets with low virality, we remove all tweets that gathered less than one retweet per hour and notes that have changed status more than 2 times. In the main specification, we focus on tweets in English but we also present our results when we use tweets in all languages in Appendix A.

We compare the diffusion of tweets with a visible Community Note rated just above the publication threshold (between 0.40 and 0.43) with tweets having no visible Note and rated just below the publication threshold (between 0.37 and 0.40). The control group is composed of 711 tweets and the treated group of 575 tweets. We consider a large number of covariates in our model to control for potential confounding factors or sources of variation. We use the log of the number of followers of the user who posted the tweet, a dummy variable to capture if the tweet contains an image, an url, a mention or a hashtag, and day-of-the week and time of the day dummies. We also add as a control the sentiment of the tweet computed using Vader \citep{hutto2014vader} and the topic of the tweet identified through a Latent Dirichlet Allocation (LDA) \citep{blei2003latent}. The nine topics that we identify with the LDA (see Appendix B) are related to the most controversial topics on social media, including politics, the Covid-19, the war between Russia and Ukraine, and the war between Israel and Hamas.

As a robustness check, we also use another methodology that matches each tweet within the treated group with a corresponding tweet from the control group based on the pre-treatment outcome in level. This methodological approach ensures that variations in tweet virality following the treatment are not influenced by pre-existing disparities, both in terms of levels and functional characteristics, prior to the intervention. We only retain matched pairs where the pre-processing diffusion process does not differ by more than 1\%. We identify a total of 1700 pairs satisfying this criterion. Subsequently, we assess the observed diffusion of treated tweets in comparison to the counterfactual diffusion of their matched counterparts, and we conduct a comparative analysis of post-treatment diffusion between the treated and control groups, employing a paired t-test.

\section{Results}
\label{sec:headings}
\begin{figure} 
    \centering
    \caption{Average Effect by length of exposure (log retweets)}
     \includegraphics[trim={0cm 0cm 0 0.7cm},clip,scale=0.65]{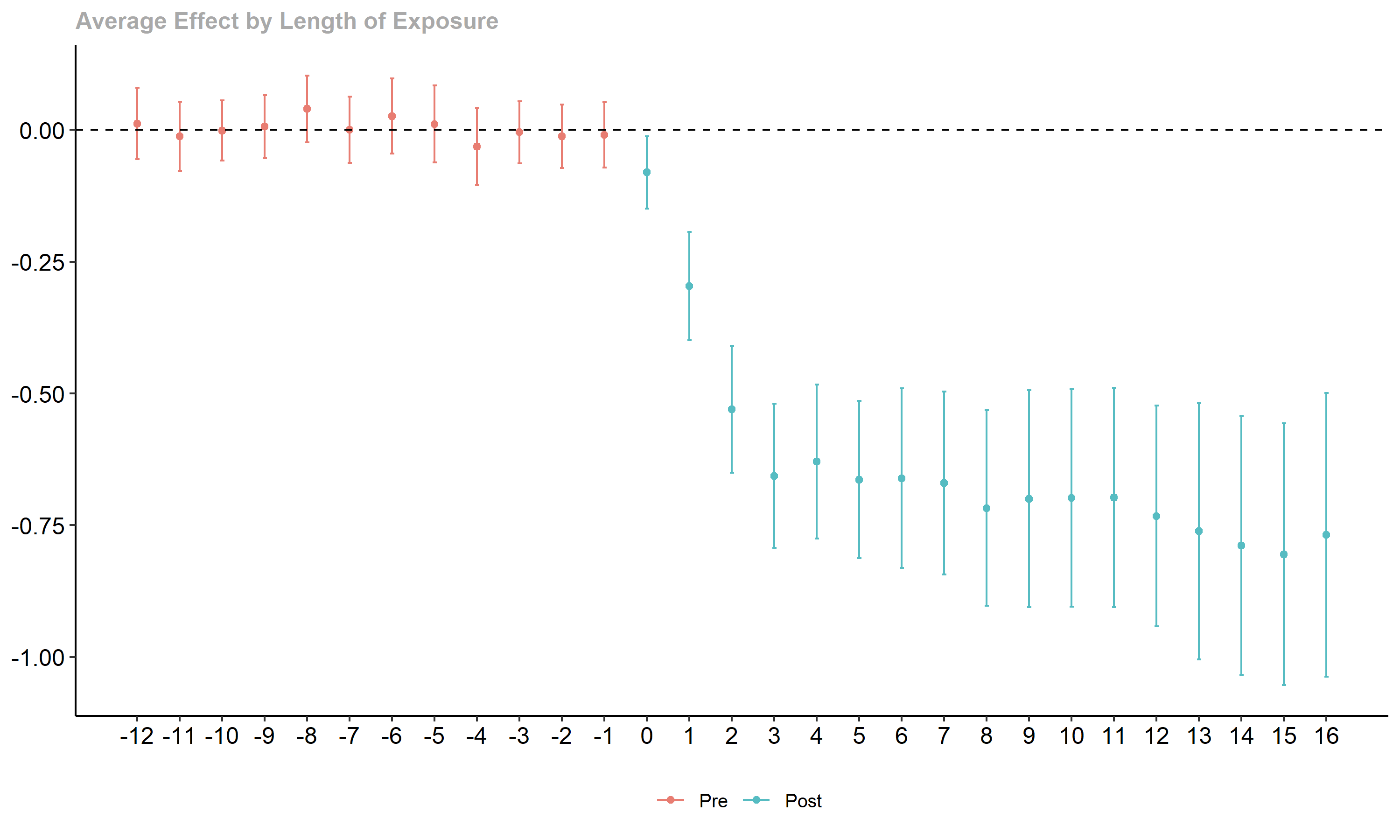} 
    \begin{center}Note : This Figure shows the average effect by length of exposure before (in red) and after (in blue) the treatment. The dependent variable is the number of retweets (in log)\end{center}
    \label{att-english}
\end{figure}

We first present our results when we use the logarithmic number of retweets as our dependent variable. Figure \ref{att-english} displays the average effect by length of exposure, spanning from 12 hours prior to 24 hours after the intervention. We find an average treatment effect on the treated tweets of -0.676. Given that our dependent variable is the logarithmic number of retweets, this translates to a 49.1\% reduction in retweet numbers (exp(-0.676) - 1). This effect is similar to the impact of warning labels \citep{pennycook2020implied} or fact-checking \citep{henry2022checking} on sharing intentions (46\%, and 45\% respectively). Additionally, we note no significant difference in the spread of tweets between the control and treated groups prior to the intervention. This observation supports the efficacy of our method, which focuses on tweets near a 0.4 threshold and accounts for various potential covariates, in establishing the causal influence of the intervention.

Our findings are consistent when we align tweets based on their outcomes prior to the intervention. Specifically, the average number of retweets after the intervention is 594, whereas we estimate it would have been 1250 without the intervention, indicating a reduction of 52.4\%. These two approaches - DiD and pre-treatment outcome matching - are distinctly different. DiD assumes parallel trends, focuses on English tweets, and considers only the covariates present at the time of the tweet's publication, excluding any post-publication data. On the other hand, pre-treatment outcome matching constructs a hypothetical post-treatment retweet count by exclusively considering the number of retweets between the publication and the treatment. In this latter approach, to ensure an adequate number of matches, we don't require the treated tweets and their counterparts to be similar in aspects such as topic, sentiment, language or the influence of the tweeting user. The fact that both methods yield comparable results provides a level of validation for our findings.

When we use the log number of replies and the log number of quotes as our dependent variable, we respectively find a reduction by 32.4\% and 34.6\% with the DiD (38.6\%  and 43\% using the pre-treatment outcome matching). However, quotes and replies can also be used to express opposite opinions or debunking, and thus might be affected differently by the treatment. Retweets normally mean alignment and lead to sharing the exact same tweet (with the Community Note if one is rated as helpful) but without adding any additional information. Thus, we believe that it represents the best estimate of the impact of Community Notes on user’s behavior.

Our results differ strongly from the findings of \cite{chuai2023roll}. The main reason for the difference is that \cite{chuai2023roll} use aggregate (ex-post) total number of likes and retweets and compare tweets that receive a Community Note with tweets that do not. As mentioned earlier, they do not observe the impact - at the tweet level - before and after the publication of a Note, and thus do not have a causal identification. 

The impact of post-treatment is significant, but its influence on the total number of retweets—and consequently on the overall spread of tweets—depends heavily on when the contextual information is made available. The quicker the intervention, the greater its overall effect. Thus, we use the earlier estimations combined with the exact timing of each Community Note's publication to determine the overall impact of the Community Notes. Our analysis suggests a 16.34\% reduction in the total number of retweets (11.75\% for replies and 16.87\% for quotes). As previously mentioned, the average time for a Community Note to be published is 15 hours, at which point a tweet has typically achieved almost 80\% of its reach. Yet, Community Notes are issued more promptly for tweets with high visibility, resulting in a more pronounced effect than merely halving the remaining average reach.

\section{Conclusion}
\label{sec:headings}

The X Community Notes system is innovative due to its open-source and crowdsourced design. The availability of high-frequency tweet sharing combined with information about the exact timing of a tweet publication, and the time at which a Note is made visible on the platform allows us - to the best of our knowledge - to conduct the first causal estimates of a content moderation system on real data from a large online platform. Our findings align with existing experimental research on warning labels and fact-checking, demonstrating that adding context to social media posts collaboratively can reduce the spread of misinformation by almost half. This effect is likely a lower bound, considering the significant increase in the number of deleted tweets for which we lack the diffusion process. Despite this substantial post-treatment effect, the overall impact on the spread of tweets is relatively modest (-16.34\% to -20.75\% for retweets). This indicates that the speed of content moderation as it currently stands may not be adequate to significantly limit the spread of misleading content on social media platforms.

\newpage

\bibliographystyle{apalike}  
\bibliography{references}

\begin{thebibliography}{}

\bibitem[Allcott and Gentzkow, 2017]{allcott2017social}
Allcott, H. and Gentzkow, M. (2017).
\newblock Social media and fake news in the 2016 election.
\newblock {\em Journal of economic perspectives}, 31(2):211--236.

\bibitem[Allen et~al., 2021]{allen2021scaling}
Allen, J., Arechar, A.~A., Pennycook, G., and Rand, D.~G. (2021).
\newblock Scaling up fact-checking using the wisdom of crowds.
\newblock {\em Science advances}, 7(36):eabf4393.

\bibitem[Blei et~al., 2003]{blei2003latent}
Blei, D.~M., Ng, A.~Y., and Jordan, M.~I. (2003).
\newblock Latent dirichlet allocation.
\newblock {\em Journal of machine Learning research}, 3(Jan):993--1022.

\bibitem[Brashier et~al., 2021]{brashier2021timing}
Brashier, N.~M., Pennycook, G., Berinsky, A.~J., and Rand, D.~G. (2021).
\newblock Timing matters when correcting fake news.
\newblock {\em Proceedings of the National Academy of Sciences}, 118(5):e2020043118.

\bibitem[Callaway and Sant’Anna, 2021]{callaway2021difference}
Callaway, B. and Sant’Anna, P.~H. (2021).
\newblock Difference-in-differences with multiple time periods.
\newblock {\em Journal of econometrics}, 225(2):200--230.

\bibitem[Chuai et~al., 2023]{chuai2023roll}
Chuai, Y., Tian, H., Pr{\"o}llochs, N., and Lenzini, G. (2023).
\newblock The roll-out of community notes did not reduce engagement with misinformation on twitter.
\newblock {\em arXiv preprint arXiv:2307.07960}.

\bibitem[Friggeri et~al., 2014]{friggeri2014rumor}
Friggeri, A., Adamic, L., Eckles, D., and Cheng, J. (2014).
\newblock Rumor cascades.
\newblock In {\em proceedings of the international AAAI conference on web and social media}, volume~8, pages 101--110.

\bibitem[Gillespie, 2018]{gillespie2018custodians}
Gillespie, T. (2018).
\newblock {\em Custodians of the Internet: Platforms, content moderation, and the hidden decisions that shape social media}.
\newblock Yale University Press.

\bibitem[Grinberg et~al., 2019]{grinberg2019fake}
Grinberg, N., Joseph, K., Friedland, L., Swire-Thompson, B., and Lazer, D. (2019).
\newblock Fake news on twitter during the 2016 us presidential election.
\newblock {\em Science}, 363(6425):374--378.

\bibitem[Henry et~al., 2022]{henry2022checking}
Henry, E., Zhuravskaya, E., and Guriev, S. (2022).
\newblock Checking and sharing alt-facts.
\newblock {\em American Economic Journal: Economic Policy}, 14(3):55--86.

\bibitem[Hutto and Gilbert, 2014]{hutto2014vader}
Hutto, C. and Gilbert, E. (2014).
\newblock Vader: A parsimonious rule-based model for sentiment analysis of social media text.
\newblock In {\em Proceedings of the international AAAI conference on web and social media}, volume~8, pages 216--225.

\bibitem[Lazer et~al., 2018]{lazer2018science}
Lazer, D.~M., Baum, M.~A., Benkler, Y., Berinsky, A.~J., Greenhill, K.~M., Menczer, F., Metzger, M.~J., Nyhan, B., Pennycook, G., Rothschild, D., et~al. (2018).
\newblock The science of fake news.
\newblock {\em Science}, 359(6380):1094--1096.

\bibitem[Martel and Rand, 2023]{martel2023misinformation}
Martel, C. and Rand, D.~G. (2023).
\newblock Misinformation warning labels are widely effective: A review of warning effects and their moderating features.
\newblock {\em Current Opinion in Psychology}, page 101710.

\bibitem[Mena, 2020]{mena2020cleaning}
Mena, P. (2020).
\newblock Cleaning up social media: The effect of warning labels on likelihood of sharing false news on facebook.
\newblock {\em Policy \& internet}, 12(2):165--183.

\bibitem[Pennycook et~al., 2020]{pennycook2020implied}
Pennycook, G., Bear, A., Collins, E.~T., and Rand, D.~G. (2020).
\newblock The implied truth effect: Attaching warnings to a subset of fake news headlines increases perceived accuracy of headlines without warnings.
\newblock {\em Management science}, 66(11):4944--4957.

\bibitem[Vosoughi et~al., 2018]{vosoughi2018spread}
Vosoughi, S., Roy, D., and Aral, S. (2018).
\newblock The spread of true and false news online.
\newblock {\em science}, 359(6380):1146--1151.

\bibitem[Wojcik et~al., 2022]{wojcik2022birdwatch}
Wojcik, S., Hilgard, S., Judd, N., Mocanu, D., Ragain, S., Hunzaker, M., Coleman, K., and Baxter, J. (2022).
\newblock Birdwatch: Crowd wisdom and bridging algorithms can inform understanding and reduce the spread of misinformation.
\newblock {\em arXiv preprint arXiv:2210.15723}.

\end{thebibliography}

\newpage
\section*{Appendix A}

\begin{figure}[h] 
    \centering
    \caption{Average Effect by length of exposure (log retweets) - All tweets}
    \includegraphics[trim={0cm 0cm 0 0.7cm},clip,scale=0.65]{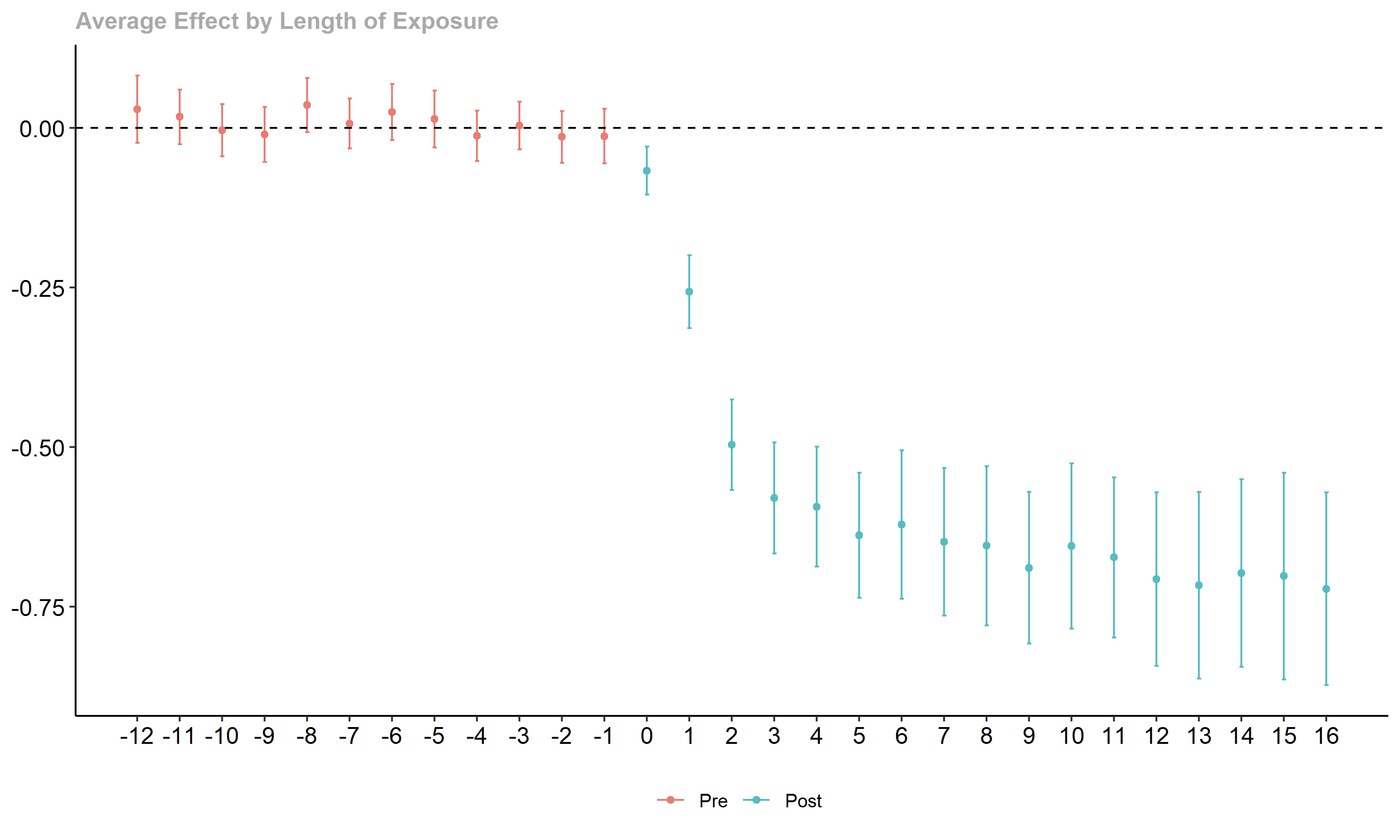} 
    \begin{center}Note : This Figure shows the average effect by length of exposure before (in red) and after (in blue) the treatment. The dependent variable is the number of retweets (in log). We consider all tweets and not only tweets in English as in Figure  \ref{att-english}. \end{center}
\end{figure}

\newpage
\section*{Appendix B}

This appendix presents the 9 topics identified through LDA across all English tweets that have received a Note.

\bigskip
\begin{figure}[h]
    \centering
    \includegraphics[trim={3cm 1cm 0 3cm},clip,scale=0.18]{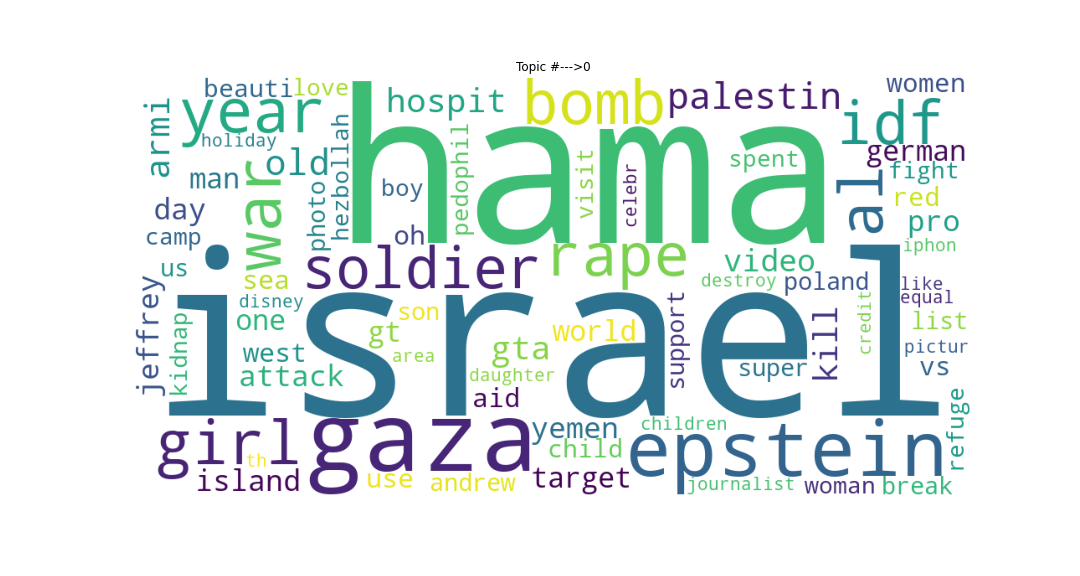}
    \includegraphics[trim={3cm 1cm 0 3cm},clip,scale=0.18]{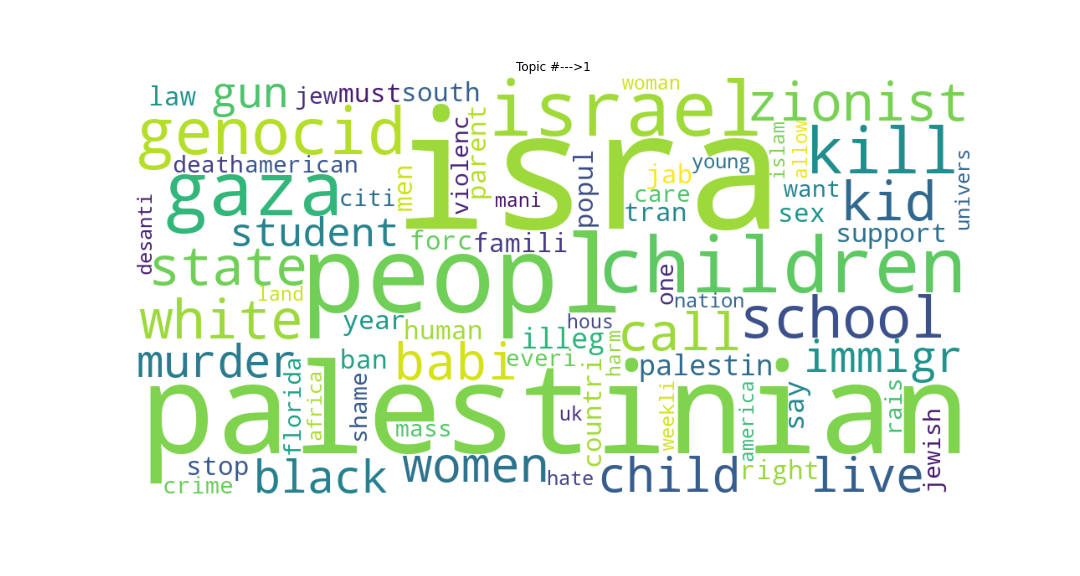}
    \includegraphics[trim={3cm 1cm 0 3cm},clip,scale=0.18]{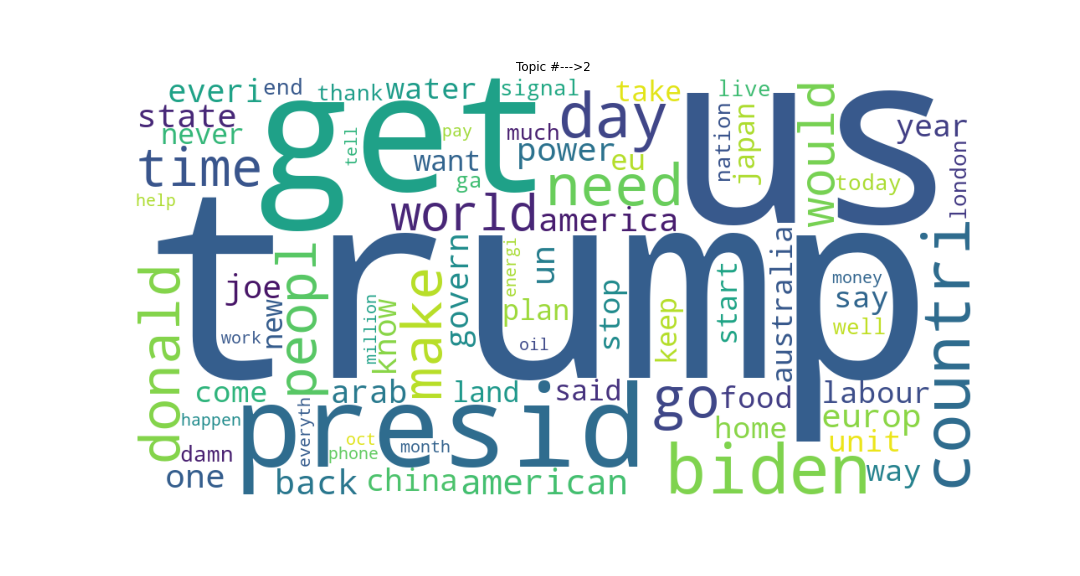}
    \includegraphics[trim={3cm 1cm 0 3cm},clip,scale=0.18]{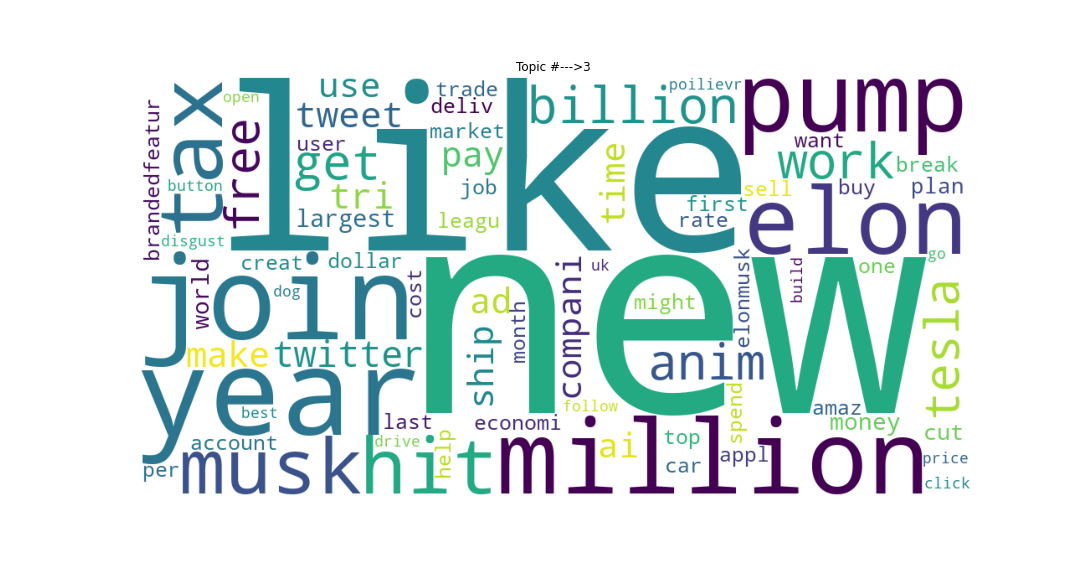}
    \includegraphics[trim={3cm 1cm 0 3cm},clip,scale=0.18]{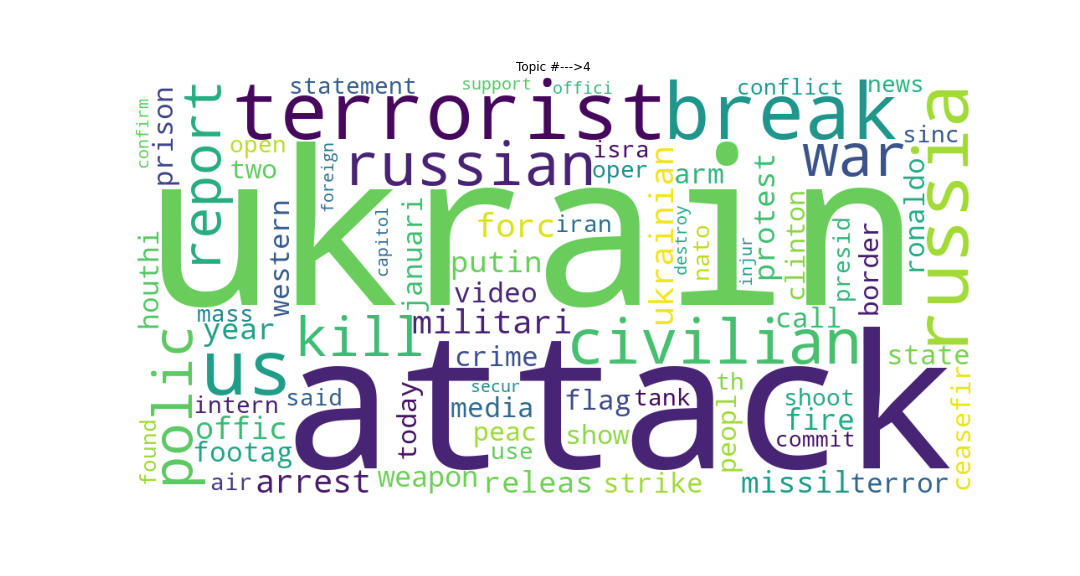}
    \includegraphics[trim={3cm 1cm 0 3cm},clip,scale=0.18]{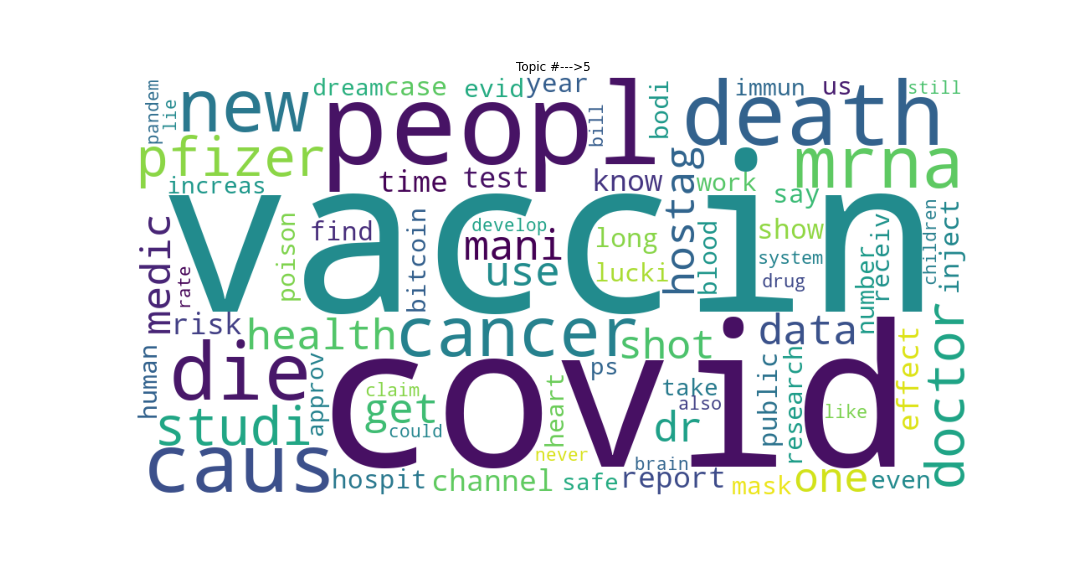}
    \includegraphics[trim={3cm 1cm 0 3cm},clip,scale=0.18]{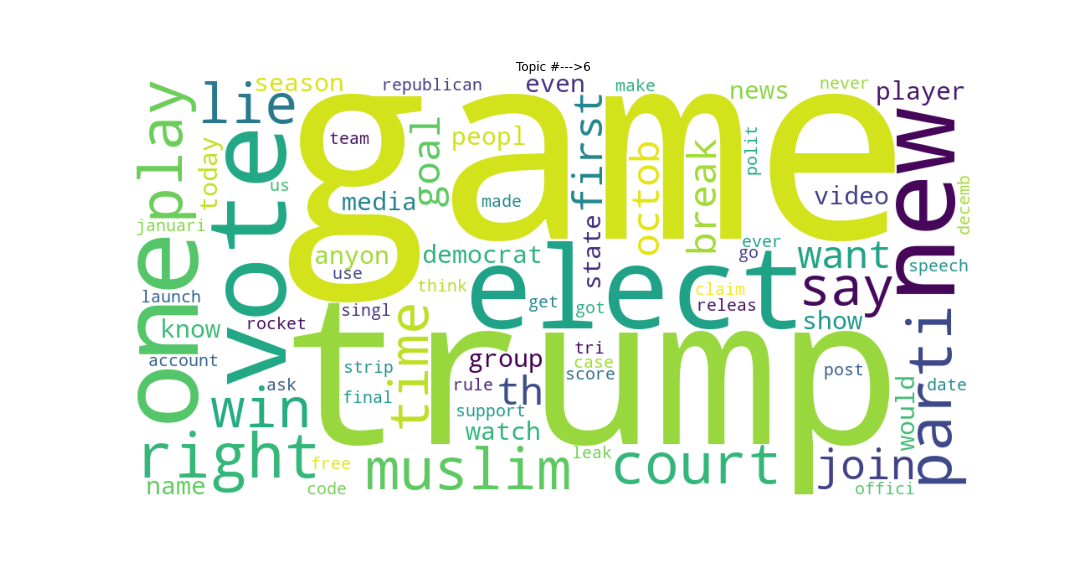}
    \includegraphics[trim={3cm 1cm 0 3cm},clip,scale=0.18]{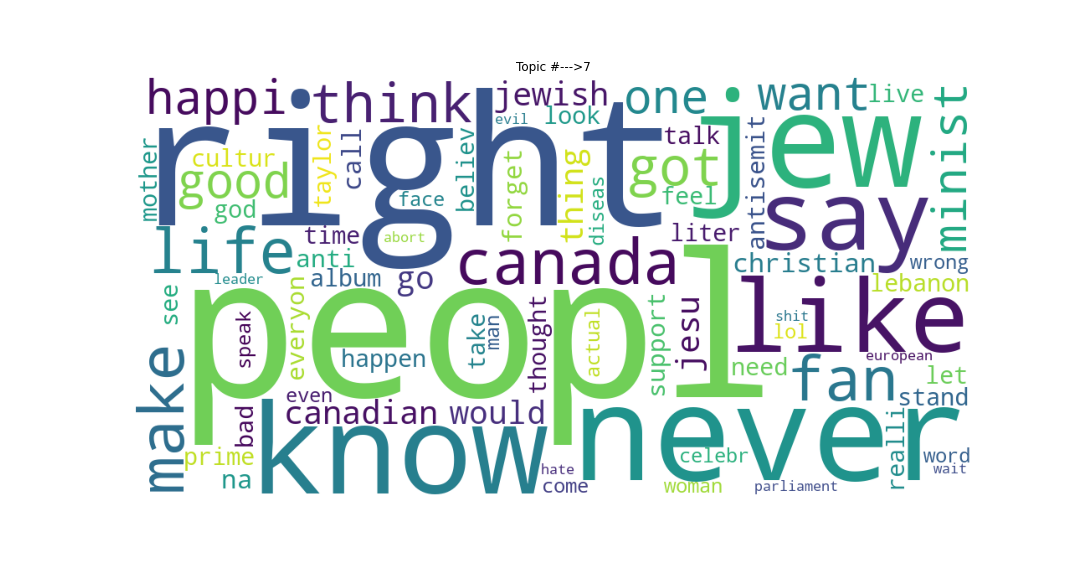}
    \includegraphics[trim={3cm 1cm 0 3cm},clip,scale=0.18]{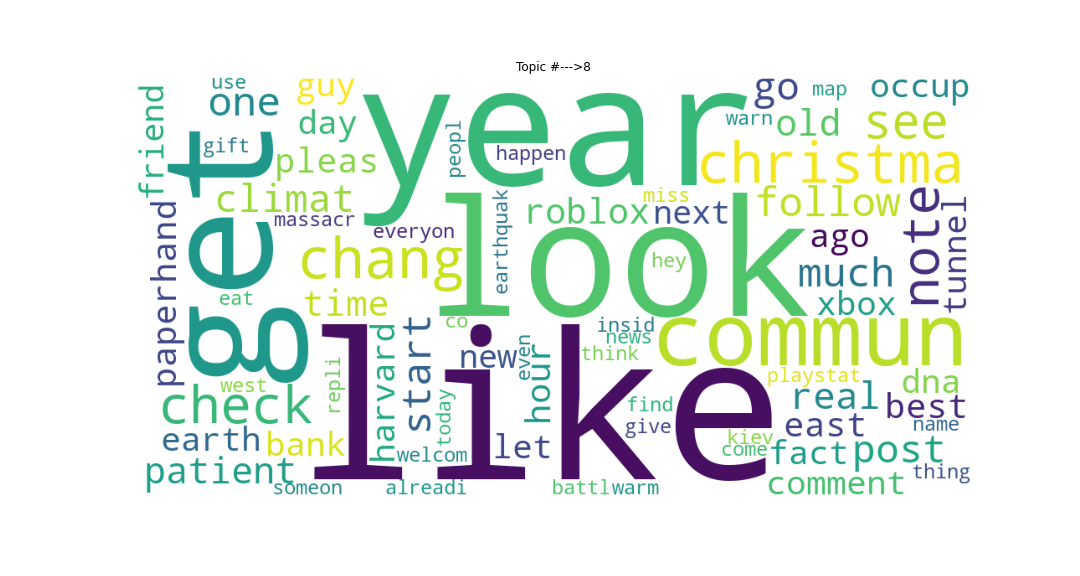}
    
\end{figure}

\end{document}